\documentclass[a4paper,10pt,twoside]{cpc-hepnp}

\usepackage{multicol}
\usepackage{graphicx}
\usepackage{booktabs}
\usepackage{amssymb,bm,mathrsfs,bbm,amscd}
\usepackage[tbtags]{amsmath}
\usepackage{lastpage}
\usepackage{CJK}

\begin{document}

\title{Beam dynamics design of the main accelerating section with KONUS in the CSR-LINAC}

\author{%
      ZHANG Xiao-Hu$^{1,2;1)}$\email{zhangxiaohu@impcas.ac.cn}%
\quad YUAN You-Jin$^1$
\quad XIA Jia-Wen$^1$ \\
\quad YIN Xue-Jun$^1$
\quad DU Heng$^{1,2}$
}
\maketitle

\address{%
$^1$ Institute of Modern Physics, Chinese Academy of Sciences, Lanzhou, 730000, China\\
$^2$ University of Chinese Academy of Sciences, Beijing, 100046, China\\
}

\begin{abstract}
  The CSR-LINAC injector has been proposed in Heavy Ion Research Facility in Lanzhou (HIRFL). The linac mainly consists of two parts, the RFQ and the IH-DTL. The KONUS (Kombinierte Null Grad Struktur) concept has been introduced into the DTL section. In this paper, the re-matching of the main accelerating section will be finished in the 3.7 MeV/u scheme and the new beam dynamics design up to 7 MeV/u will be also shown. Through the beam re-matching, the
  relative emittance growth has been suppressed greatly along the linac.
\end{abstract}



\begin{multicols}{2}

\section{Introduction}

 The Heavy Ion Research Facility in Lanzhou (HIRFL) has been upgraded with the HIRFL-CSR project at the end of the year 2007 and supplies 7000 hours operation time annually [1]. The injector of CSR consists of two cyclotrons, Sector Focusing Cyclotron (SFC) and Separator Sector Cyclotron (SSC). However, the linear accelerator becomes very popular as the injector of the subsequent accelerator in recent years, such as, GSI-HIS, TRIUMF-ISAC, CERN-LINAC3, HIMAC and HIT. Due to larger beam acceptance, higher transmission and higher accelerating gradient, the linac injector can supply higher intensity and better beam quality. As the number of the applicants increasing rapidly and the experimental requirement improved simultaneously, one new injector become most essential for HIRFL-CSR. The new injector called CSR-LINAC has been proposed, which will achieve the double-terminal operation in parallel and make the operation time to increase by more than 5000 hours per year and attract more comprehensive physical experiments, as shown in Fig.1.

 The CSR-LINAC injector can supply all kinds of heavy ion beam with 7 MeV/u for HIRFL-CSR. The charge-to-mass ratio is from 1/7 to 1/3 and the designed beam intensity is chosen to 3 emA. Both RFQ and DTL are essential in the whole linac scheme and the layout of this linac injector is shown in Fig.2. The beam will be accelerated to 300 keV/u in RFQ and then transported to 7 MeV/u in the main accelerating section. The main parameters of the CSR-LINAC are summarized in Table 1. The KONUS concept is introduced into the main accelerating section in order to get higher accelerating gradient. The physics design of the 3.7 MeV/u scheme has been proposed by the Institute for Applied Physics (IAP) [2]. So only the beam dynamics design of the downstream section become one new objective. In this paper, the re-matching of the 3.7 MeV/u main accelerating section is firstly finished and then the 3.7 MeV/u to 7 MeV/u beam dynamics scheme is shown completely.

\begin{center}
\includegraphics[width=7cm,height=6cm]{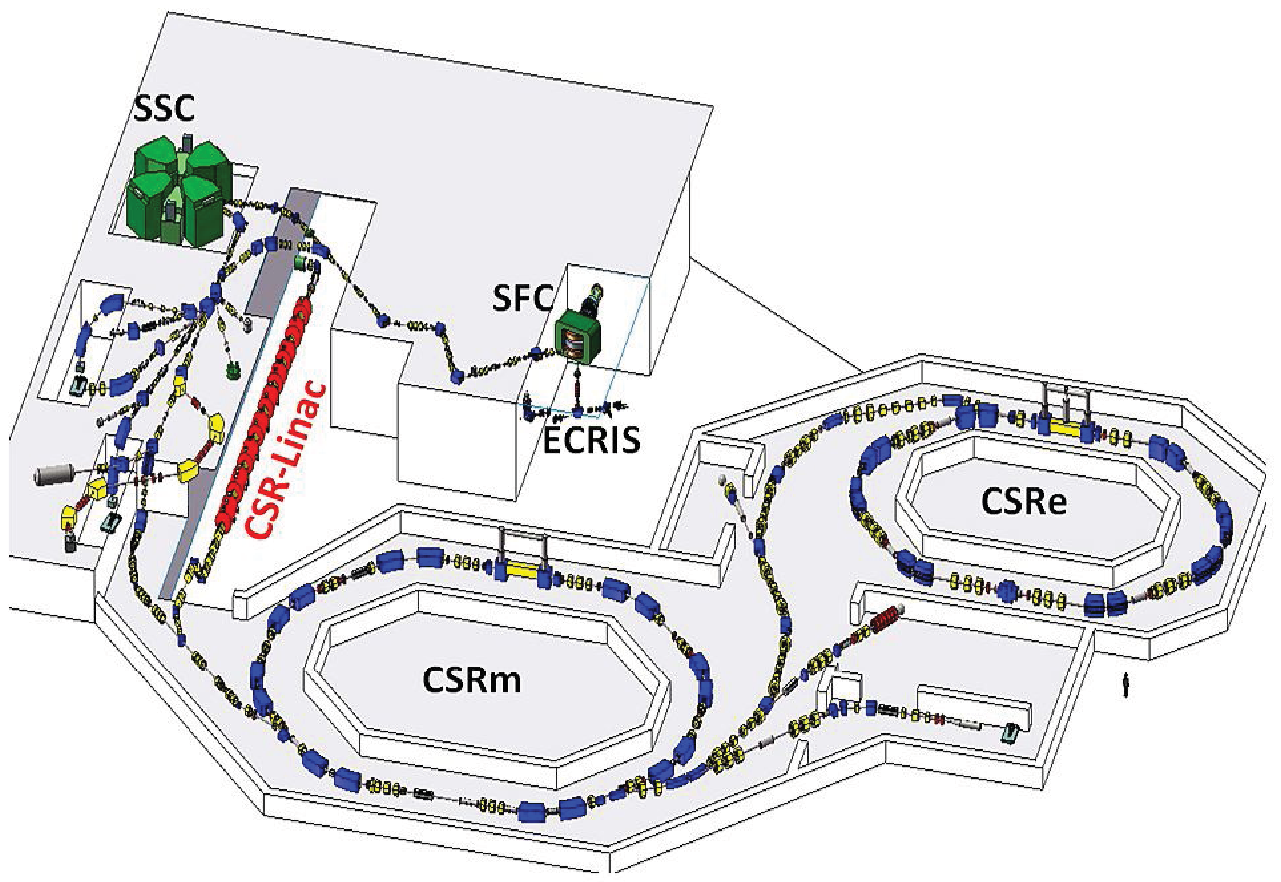}
\figcaption{\label{fig1} The layout of HIRFL-CSR  with the new injector.}
\end{center}

\begin{center}
\tabcaption{\label{tab1}  Main Parameters of the CSR-Linac.}
\footnotesize
\begin{tabular*}{80mm}{c@{\extracolsep{\fill}}ccc}
\toprule
Q/A & 1/3 - 1/7 & - \\ \hline
Emittance(norm, 90\%)  & 0.4  & pi.mm.mrad \\ \hline
Frequency  & 108.48/216.96 & MHz \\ \hline
Beam current  & 3 & emA \\ \hline
Duration  & 3 & ms \\ \hline
Repetition & 10  & Hz \\ \hline
RFQ input/output energy & 4/300 & keV/u \\ \hline
DTL input/output energy & 0.3/7 & MeV/u \\ \hline
Transmission(design) & 90 & \% \\
\bottomrule
\end{tabular*}
\end{center}

\end{multicols}
\begin{center}
\includegraphics[width=14cm,height=2cm]{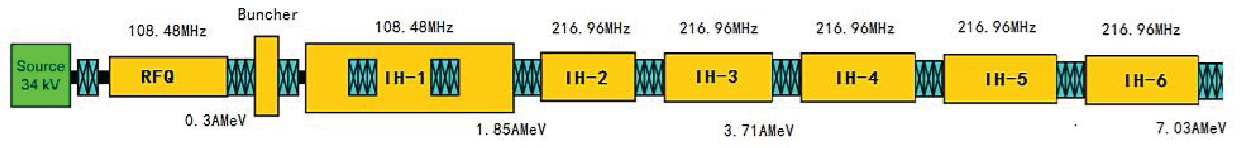}
\figcaption{\label{fig2} Preliminary layout of the 7 MeV/u CSR-LINAC proposal.}
\end{center}

\begin{multicols}{2}

\section{Re-matching of the 3.7 MeV/u scheme}

  The 3.7 MeV/u scheme of CSR-LINAC had been proposed four years ago. The LORASR code is applied to the beam dynamics of the KONUS concept in the main accelerating section. However, this scheme can be optimized any more to get better quality beam and larger error tolerance. As seen from Fig.3(left), the beam matching is not good along the whole DTL section, especially in the first DTL cavity. The non-symmetric beam will cause the emittance growth and beam coupling in the symmetric RF electric
  field. For matching from the exit of RFQ to the DTL section easily, the 5-Quadrupole scheme will be replaced by the 6-Quadrupole scheme in the MEBT section. The
  symmetric beam matching method is adopted to reduce the beam coupling in the RF field. In this case, The emittance growth evolution is shown in Fig.4. After the re-matching, the relative RMS emittance growth is reduced greatly in all three phase space. The maximum envelope is decreased by 3 mm, which is benefit for alignment and suppressing the beam nonlinear effect, as shown in Fig.3.

\section{Up to 7 MeV/u beam dynamics}

The KONUS concept is applied for the 3.7 to 7 MeV/u beam dynamics scheme. LORASR is used as the only code for the KONUS concept [3]. The period structure concept is proposed in the KONUS beam dynamics design. A KONUS period is composed of three sections with separated function. The first section consists of a few gaps with a negative synchronous phase of typically from -25¡ãto -35¡ãand acts as a rebuncher. Then the beam is injected into the main accelerating section with surplus energy and phase compared with a synchronous particle. Finally, the multi-gap section is followed by the transverse focusing elements, such as the magnetic quadrupole triplets[4].

In the beam dynamics design of the KONUS period structure, it is very important to choose the key parameters [5], such as:
  \begin{itemize}
  \item The effective voltage distribution per section
  \item The radius and gap length per cell
  \item The gap number of the rebuncher section
  \item The gap number of the 0¡ãsection
  \item The starting phase and energy of the 0¡ãsection\\
  \end{itemize}
\end{multicols}
\begin{multicols}{2}
\begin{flushright}
\includegraphics[width=8cm,height=2.8cm]{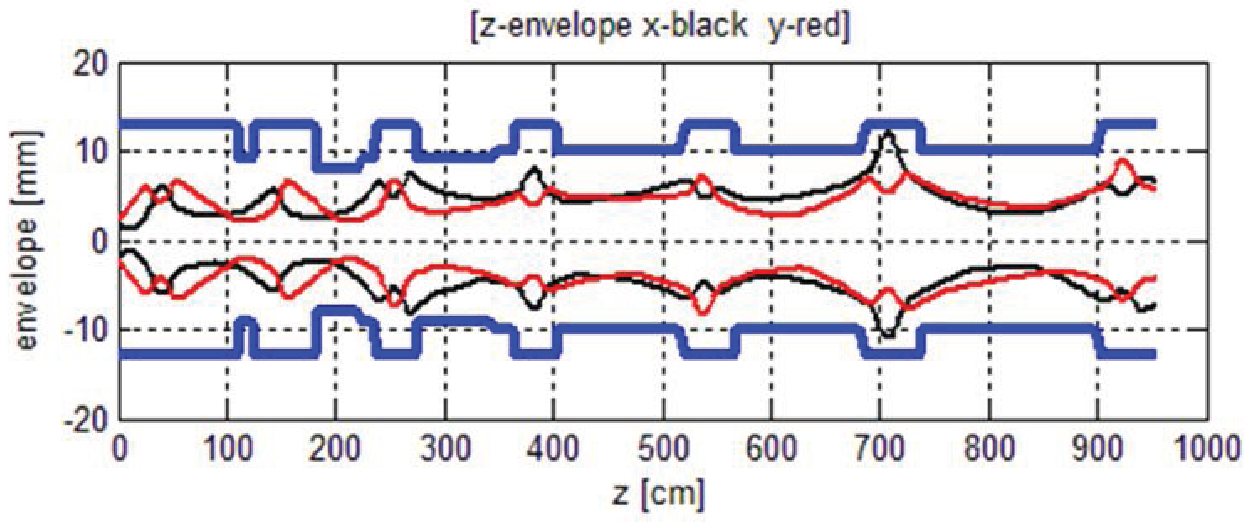}
\end{flushright}
\begin{flushleft}
\includegraphics[width=8cm,height=2.8cm]{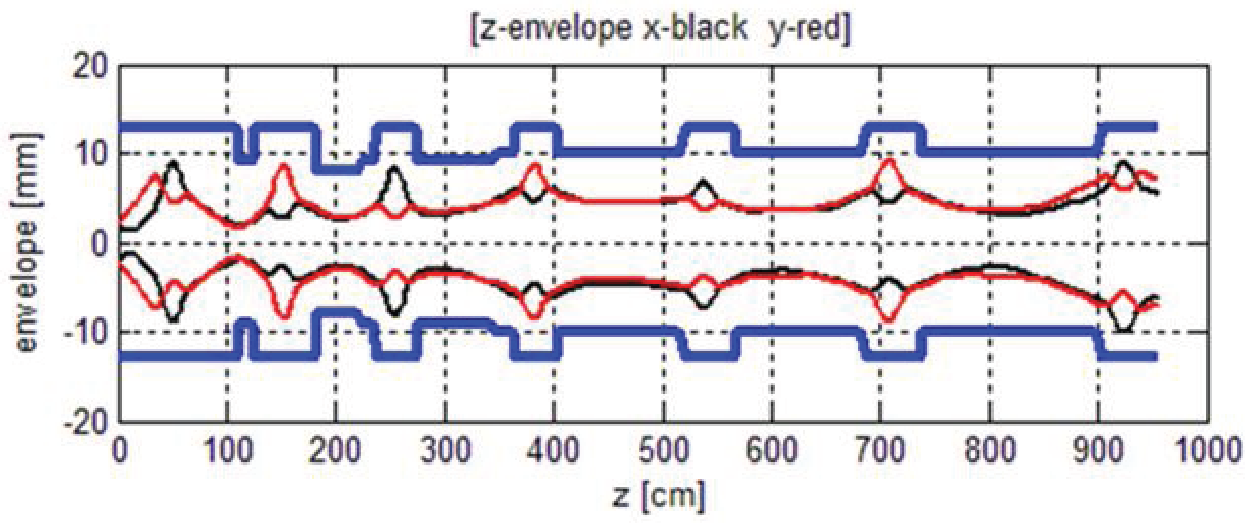}
\end{flushleft}
\end{multicols}
\figcaption{\label{fig3} The beam envelope as the function of
the position z before (left) and after (right) the beam
re-matching.}

\begin{multicols}{2}
\begin{flushright}
\includegraphics[width=8cm,height=3.3cm]{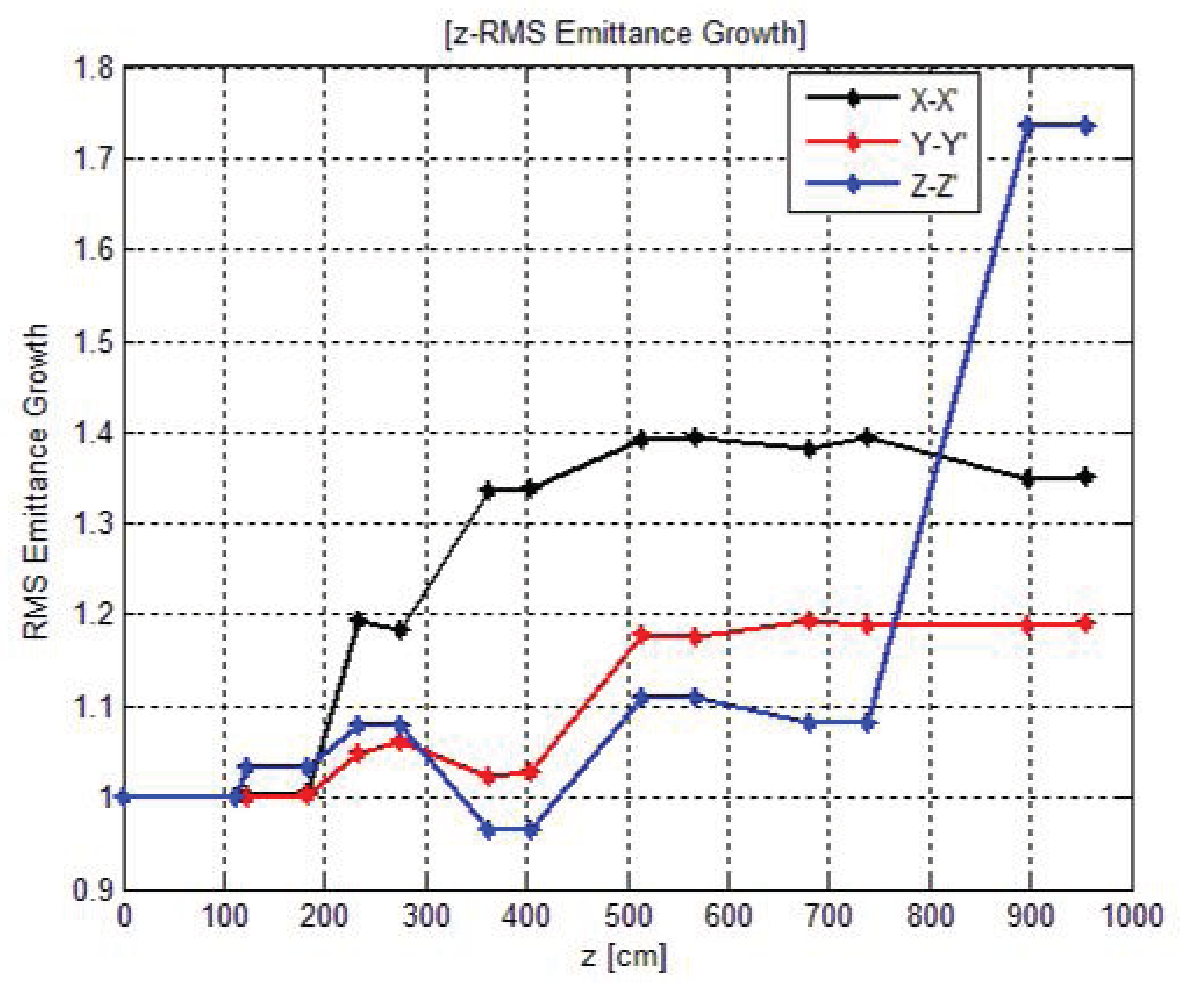}
\end{flushright}
\begin{flushleft}
\includegraphics[width=8cm,height=3.3cm]{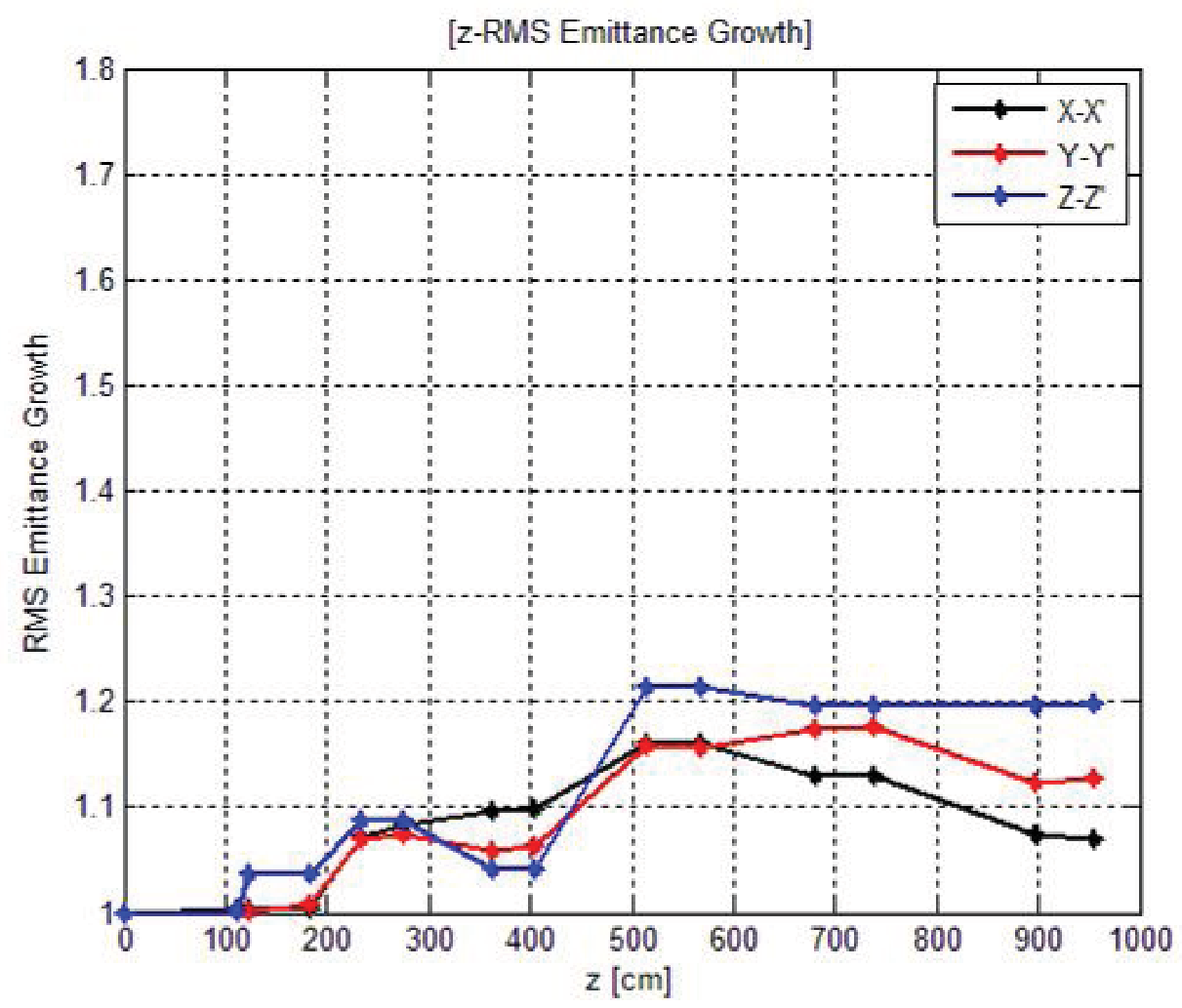}
\end{flushleft}
\end{multicols}

\figcaption{\label{fig4} The relative emittance growth as
the function of the position z before (left) and after
(right) the beam re-matching.}

\begin{multicols}{2}
\begin{flushright}
\includegraphics[width=8cm,height=4.5cm]{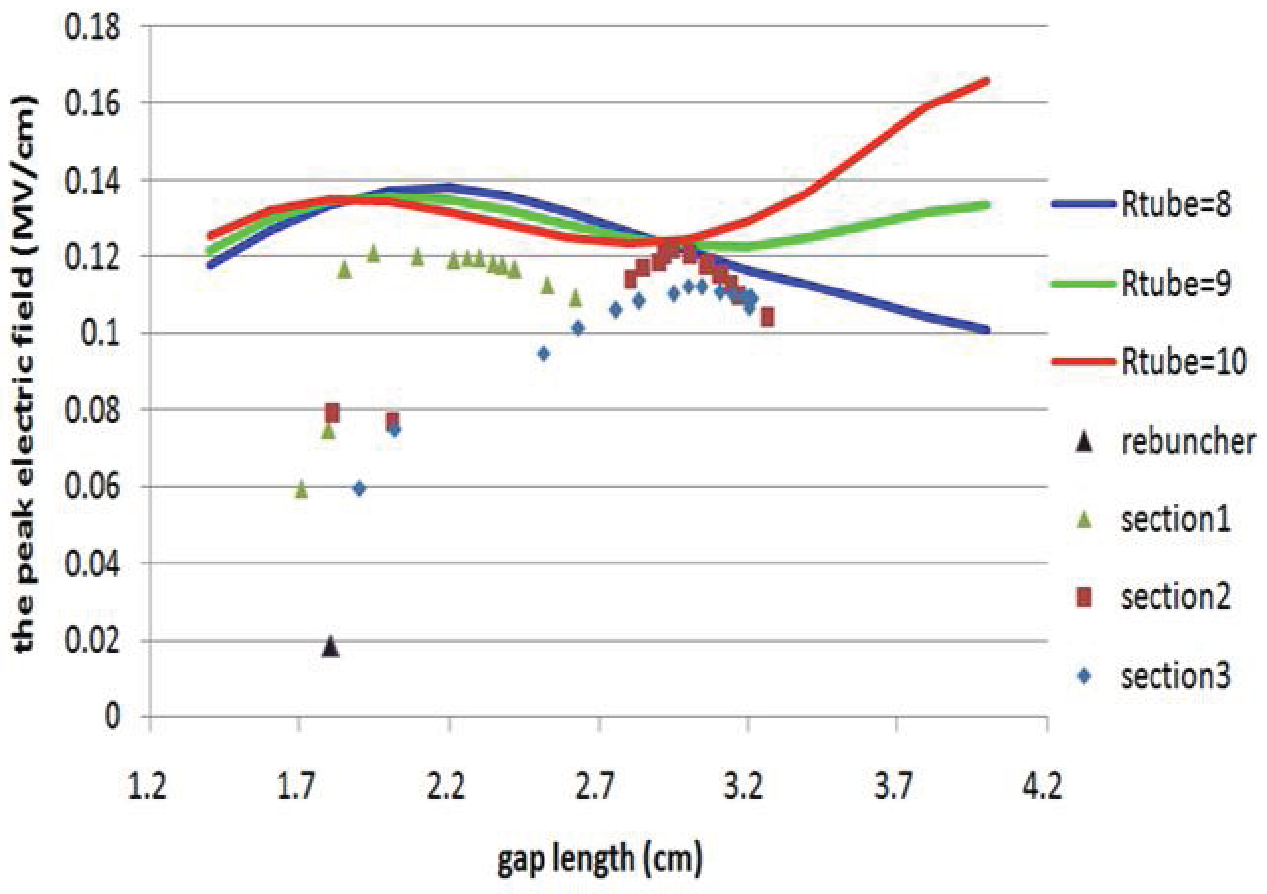}
\end{flushright}
\begin{flushleft}
\includegraphics[width=8cm,height=4.5cm]{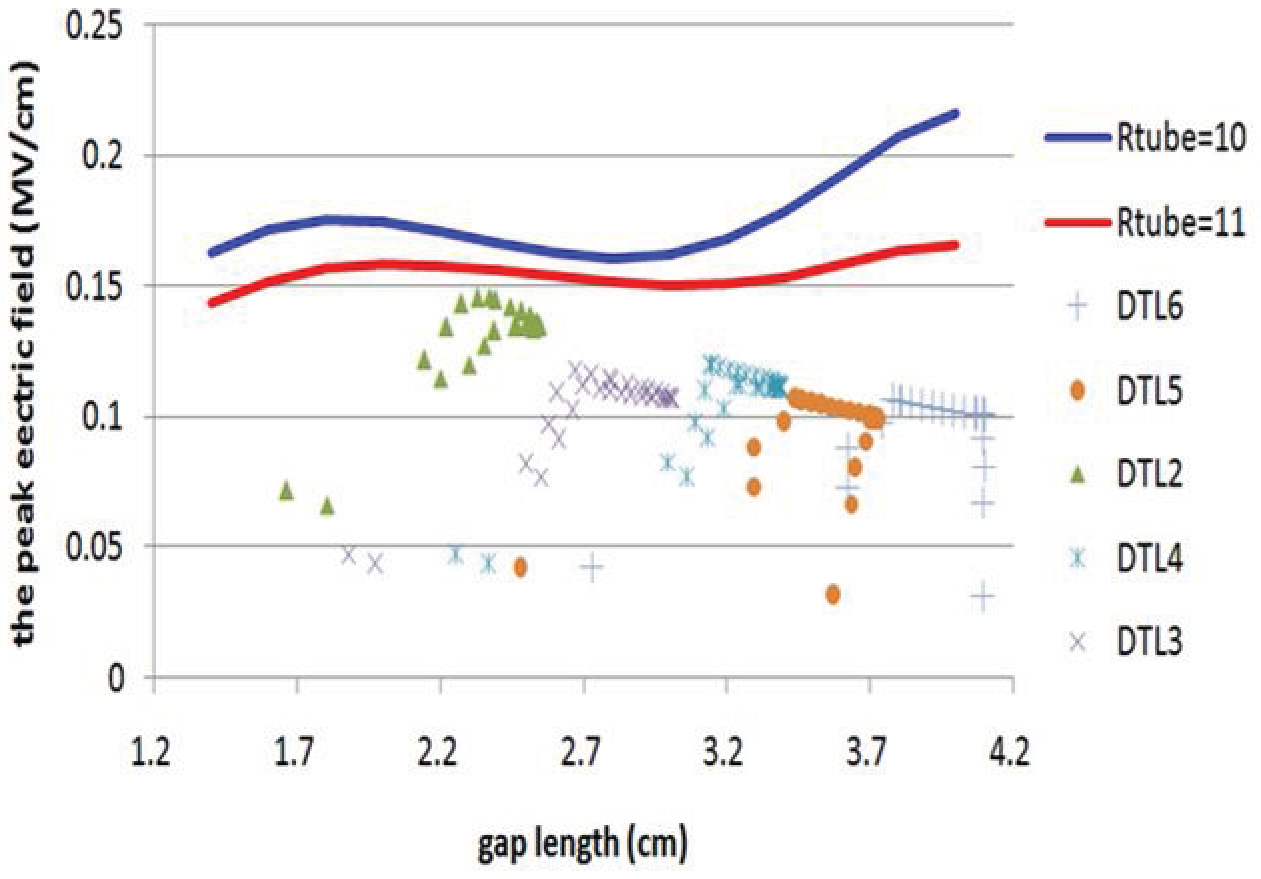}
\end{flushleft}
\end{multicols}

\figcaption{\label{fig5} The dots stand for the peak
electric field distribution per cavity at the frequency
of 108.48 MHz (right) and 216.96 MHz (left). The lines
mean the maximum peak electric field corresponding to the
spark as the function of the gap length, at the different
tube radius.}

\begin{multicols}{2}

\subsection{The choice of the effective voltage distribution}

The effective voltage per gap should be chosen to ensure that the spark don't appear during the commissioning and operation. The effective voltage per cell mainly depends on the operation frequency, the tube radius, the gap length and the geometry of the pole.The operation frequency is 108.48 MHz and 216.96 MHz in the main accelerating section, which corresponding to the spark electric field of 21.05 MV/m and 27.37 MV/m respectively, as 1.8 times of the Kilpatrick electric field is chosen. The CST STUDIO software is applied to research the relation between the peak electric field (Ep) and the maximum surface electric field (Es,max). The optimized ratio Ep/Es,max depends on the tube radius, the gap length and the geometry of the pole. So the maximum peak electric field corresponding to the spark can be given at the different tube radius and gap length, as shown in Fig.5(line)and the peak electric field distribution per DTL section is also exhibited in the Fig.5(dot). The effective voltage should be chosen in which the peak electric field per cell is below the maximum peak electric field to avoid the spark on the pole. As can be seen in Fig.5, the choice of the effective voltage distribution per section is reasonable in the 3.7 MeV/u scheme, which will verify the validity of this research once more. According to this Ep-Es,max database, the effective voltage distribution can be also chosen in the last three cavities.

\subsection{The choice of the gap number per section}
The rebuncher section is used to bunch the beam in the longitudinal phase space. Generally, the synchronous phase is chosen to -35¡ãand the gap number of this section depends on the starting focusing status in the transverse and longitudinal phase space. At the end of the rebuncher section, the beam should be focused simultaneously at three phase space so that transported through the 0¡ãsection effectively. In order to reduce the gap number of the rebuncher section, the phase spread is set to
about 45¡ãand the relative energy spread is chosen to around 1.5\%. The cell number of the rebuncher section is set to 10, 5 and 8 in DTL4, DTL5 and DTL6 separately.

The choice of the starting phase and energy depends on the phase space distribution at the entrance of the  0¡ãsection. A good choice will bring smaller emittance growth and reasonable output phase space distribution, which is benefit for the beam transport at the downstream DTL section. In addition, the simultaneous matching is an important criterion at the transverse and longitudinal phase space, which can determine the gap number of the 0¡ãsection. The cell number of the  0¡ãsection is set to 24, 24 and 16 in DTL4, DTL5 and DTL6 separately.

The evolution of the reference particle in the longitudinal phase space is shown in Fig.6. As can be seen, the trajectory of the center particle is anomalous in the longitudinal phase space of the fourth DTL. Because no consideration is shown on updating to higher energy when the last DTL is designed, the endmost beam is difficult to matching a new KONUS period in the 3.7 MeV/u scheme. The modification of the beam dynamics in the third DTL may be a good choice in the future.

\begin{center}
\includegraphics[width=7cm,height=4cm]{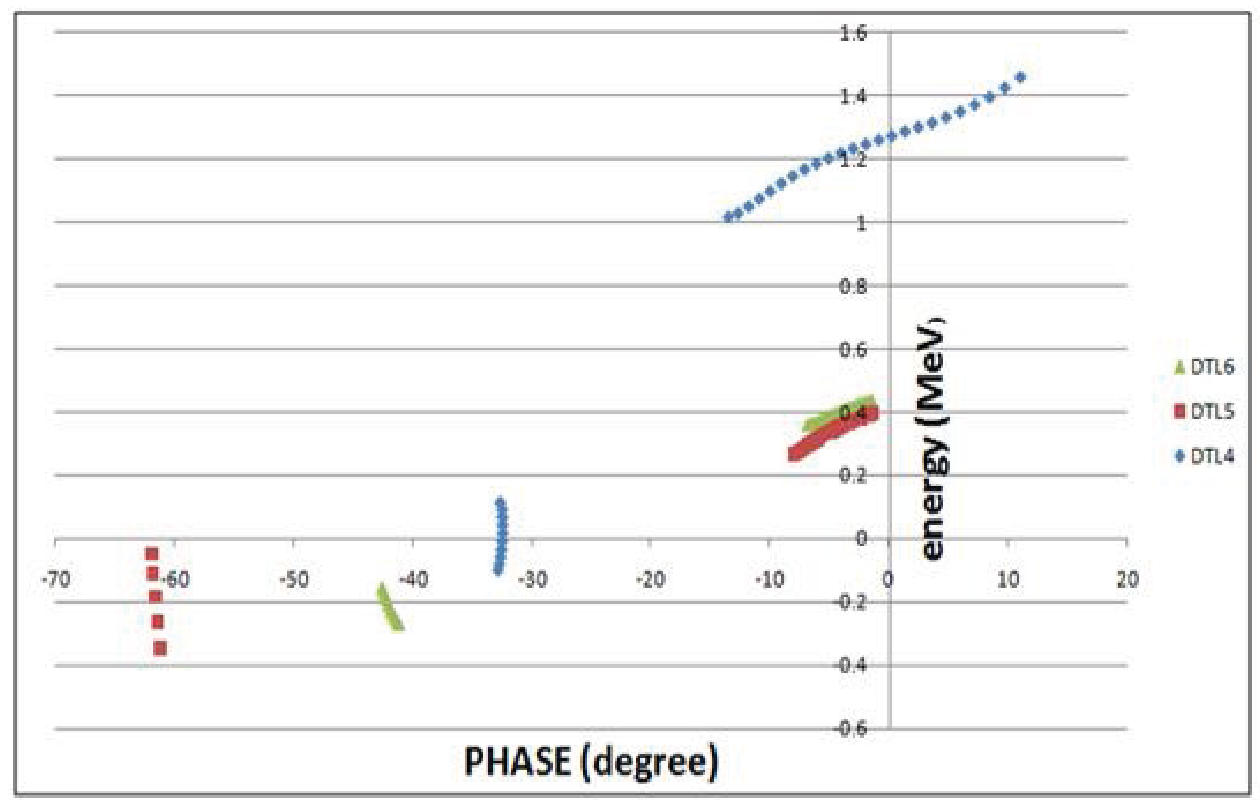}
\figcaption{\label{fig6} The evolution of the reference particle
in the longitudinal phase space}
\end{center}

The aperture of the tube is chosen to 22 mm and the beam pipe in the triplet section is 26 mm, which is helpful for controlling the non-linear effect caused by the RF electric field. The gap length distribution per section will determine the peak electric field distribution and the transfer time factor. The gap length distribution per section is checked firstly to ensure the transit time factor above 0.8. The peak electric field distribution along the main accelerating section is shown in Fig.7.

\begin{center}
\includegraphics[width=7cm,height=4.5cm]{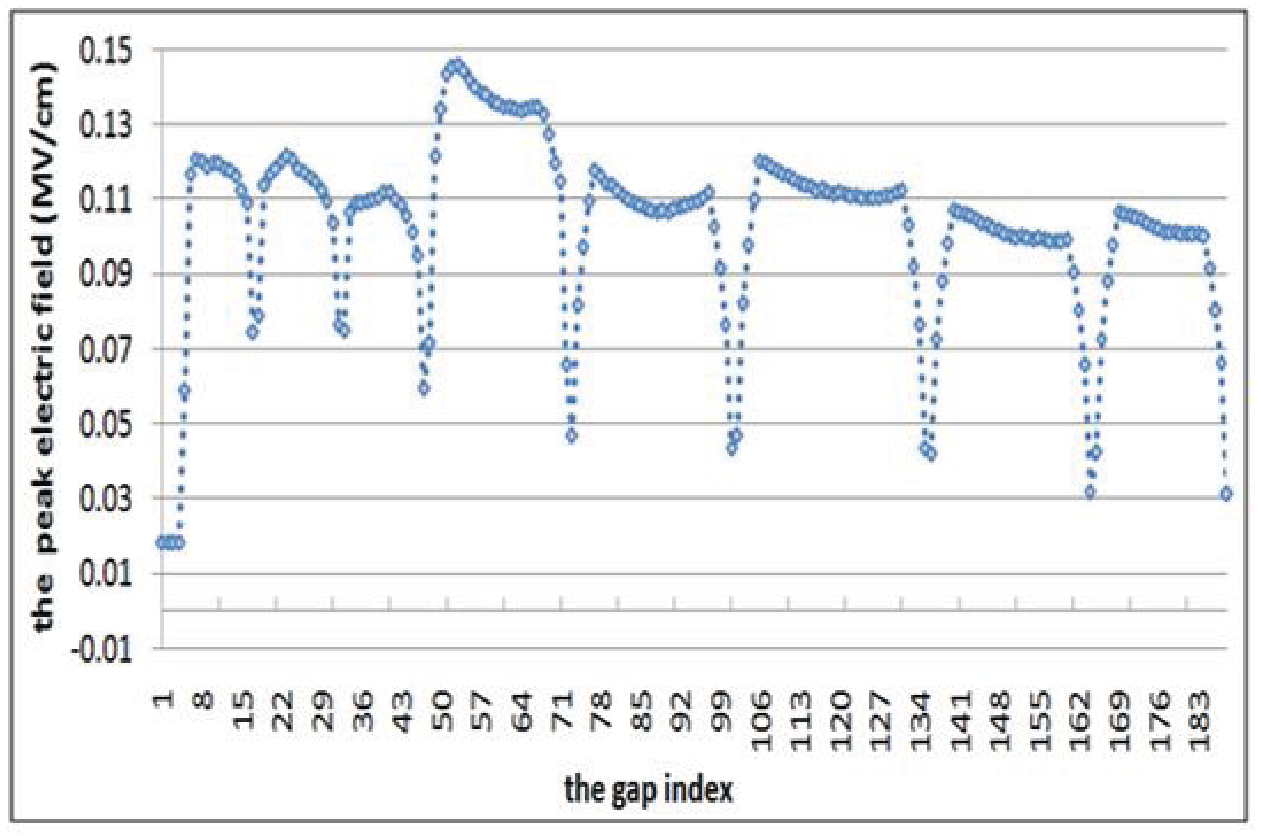}
\figcaption{\label{fig7} The peak electric field per gap as the function
 of the gap index}
\end{center}

In this design, the maximum quadrupole gradient will reach 90 T/m, which approaches the limitation for the conventional magnetic quadrupole, according to the present status in IMP.

\section{End-to-end beam dynamics}
The end-to-end beam dynamics simulation has been accomplished. The beam envelope evolution is shown in Fig.8. As can be seen, there is a good beam matching and small envelope along the linac. The maximum envelope appears at the end of DTL3 and the beam envelope be maintained down to 90\% of the tube radius. Figure 9 exhibits the phase space distributions in the input and output end of the DTL section. The distribution in the entrance is rebuilt from the phase space in the exit of RFQ. As can be seen from the distribution in the exit, the longitudinal phase space has some filament caused by the non-linear field, which will result in the longitudinal emittance growth. As exhibited in Fig.10, the relative RMS emittance growth is 6.3\%, 11.7\% for x-x', y-y' phase space, and the emittance growth reaches 28.1\%
in the longitudinal phase space.
\\
\end{multicols}
\begin{center}
\includegraphics[width=15cm,height=4cm]{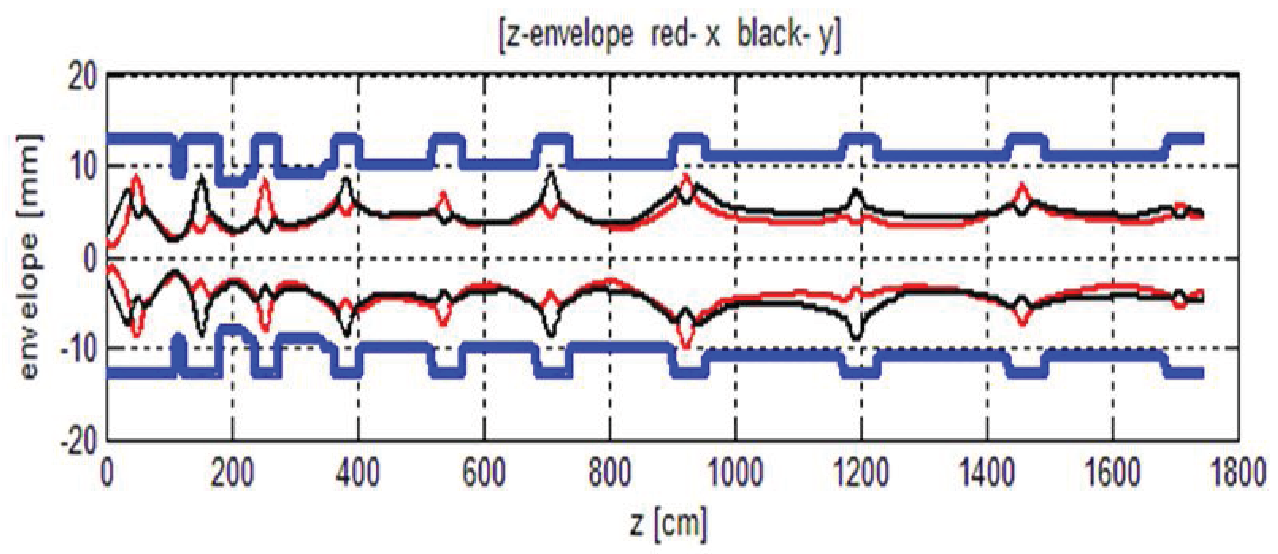}
\figcaption{\label{fig8} The x (red) and y (black) transverse envelope
as a function of position z.}
\end{center}

\begin{multicols}{2}

\begin{center}
\includegraphics[width=7cm,height=4cm]{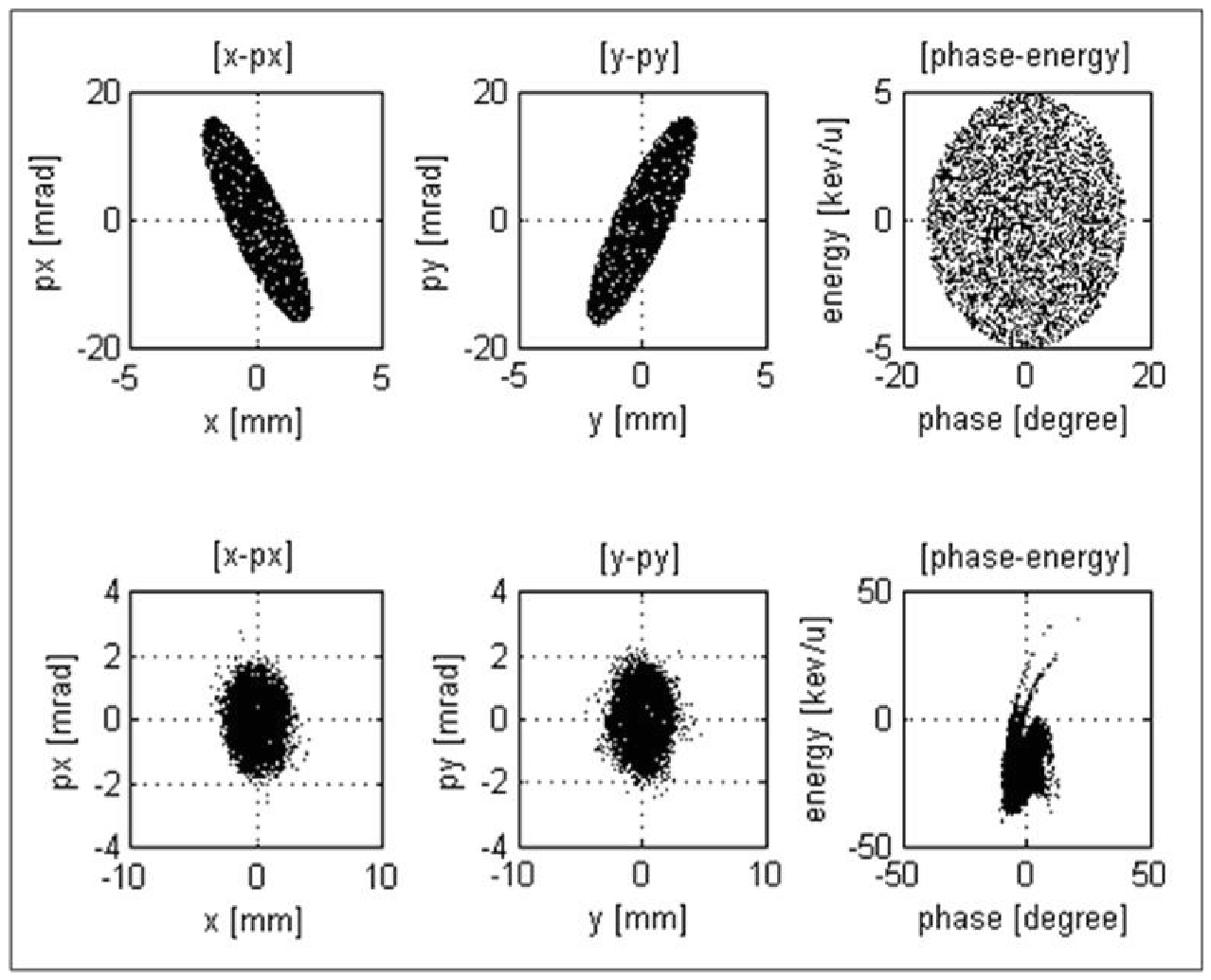}
\figcaption{\label{fig9} The phase space distribution at the
entrance (upper) and exit (lower) of the DTL section}
\end{center}

\begin{center}
\includegraphics[width=7cm,height=4cm]{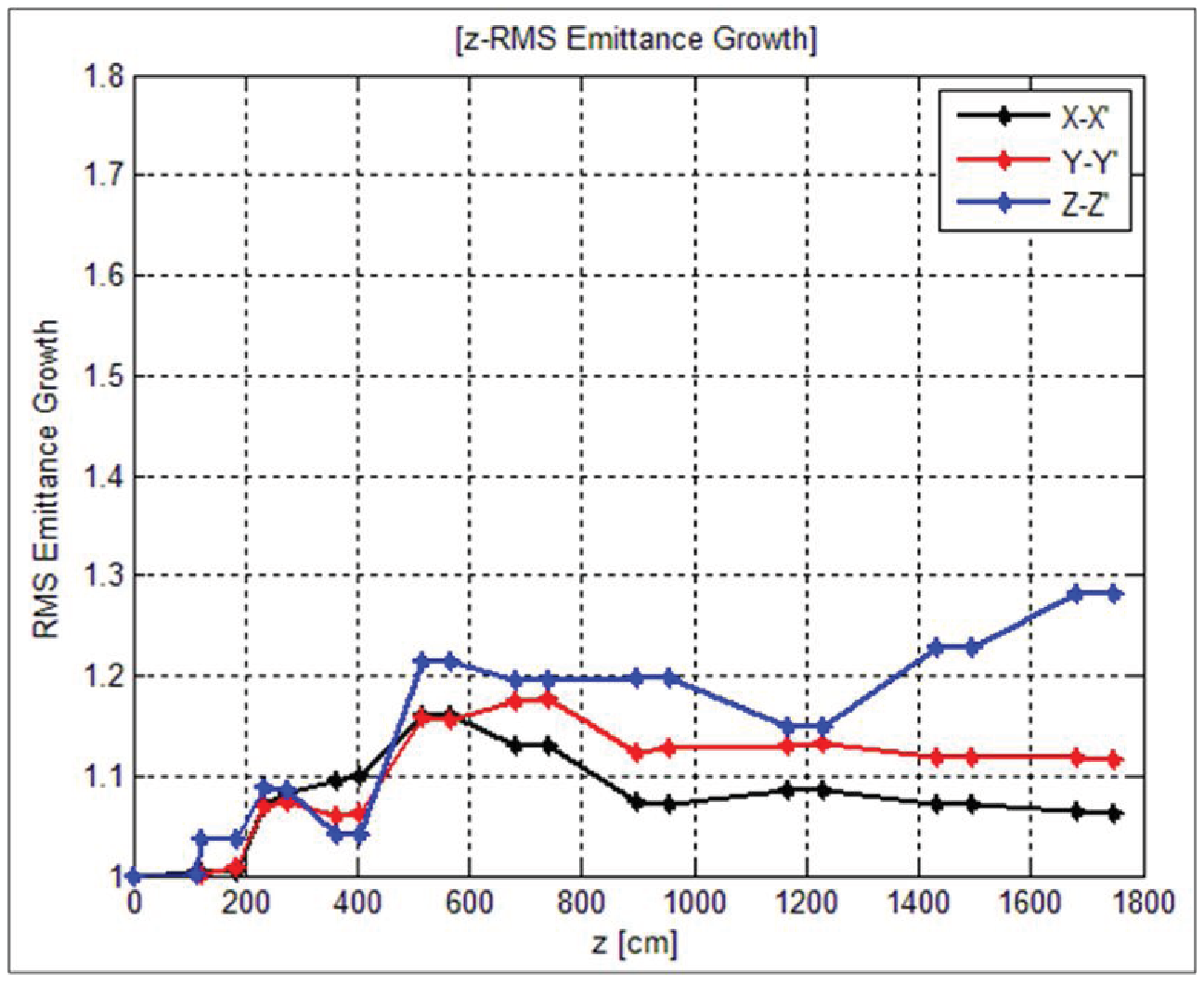}
\figcaption{\label{fig10} The relative emittance growth as a
function of position z.}
\end{center}

\end{multicols}
\begin{multicols}{2}

\section{Conclusion}
The beam re-matching of the 3.7 MeV/u beam dynamics scheme is proved advantageous for reducing the RMS emittance growth. It would be a good idea to choose the symmetric beam during the transportation in the RF electric field. The reasonable effective voltage can be chosen and this method is proved valid through the 3.7 MeV/u beam
dynamics design proposed by IAP. The 3.7 to 7 MeV/u scheme which uses only 3 cavities reveals that the KONUS structure has a high accelerating gradient. The end-to-end simulation shows that the whole beam dynamics design is reasonable in the case of no changes in the 3.7 MeV/u scheme.

\section{Acknowledgements}
The work was supported by the exports from the Frankfurt University in German. We want to express special thanks to H. Podlech, U. Ratzinger and A. Schempp (IAP) for having proposed the 3.7MeV/u heavy ion linac injector scheme for HIRFL-CSRm.

\end{multicols}

\vspace{10mm}

\vspace{-1mm}
\centerline{\rule{80mm}{0.1pt}}
\vspace{2mm}

\begin{multicols}{2}

\end{multicols}

\clearpage


\begin{thebibliography}{90}

\vspace{3mm}

\bibitem{lab1}J.W. Xia, W.L. Zhan, Y.J. Yuan, Y. Liu, X.D. Yang, Y.He, J. C. Yang and CSR group, Construction and Commissioning of the HIRFL-CSR, APAC 2007.

\bibitem{lab2} H. Podlech, U. Ratzinger, A. Schempp, IAP, J.W. Goethe University Frankfurt/Main, Design Study of A 3.7 AMeV Linac for A/Q Up To 8.5, 2009.

\bibitem{lab3} R. Tiede, G. Clemente, H. Podlech, U. Ratzinger, A. Sauer, S. Minaev, ¡®LORASR Code Development¡¯, Proc. of the 2006 EPAC Conf., Edinburgh, pp. 2194-
2196.

\bibitem{lab4} Thomas P. Wangler, RF Linear Accelerators (the 2nd edition), 2008.

\bibitem{lab5} R. Tiede, U. Ratzinger, H. Podlech, C. Zhang, G. Clemente, KONUS beam dynamics designs using H-mode cavities, Hadron Beam. 01/2008

\end{thebibliography}
\end{document}